\documentclass[reqno]{amsart}
\usepackage{graphics,amsmath}
\usepackage{t1enc}              
\usepackage[utf8]{inputenc}    

\newcount\n  \newdimen\w

\def\Repeat#1#2{\n=#1\relax\loop\ifnum       
  \n>0\relax #2\advance\n by-1\repeat}

\long\def\OMIT#1{\relax }  

\def\re#1{(\ref{#1})}   
  
\def\eqn#1#2{ \begin{align} \label{#1}         #2 \end{align}}

\def\nl#1{          \\ \label{#1}        }  

\def\delim#1#2#3{\csname\ifcase#1 relax\or   
   big\or Big\or bigg\or Bigg\fi\endcsname   
  {\ifcase#2\or\Delim#3\or\deliM#3\fi}}      
\def\Delim#1{\ifcase#1\relax\or(\or[\or\{\or<\or\langle\or|\or\|\or---{ }\fi}
\def\deliM#1{\ifcase#1\relax\or)\or]\or\}\or>\or\rangle\or|\or\|\or{ }---\fi}

\def\largerfrac#1#2#3{      
  \whichtypesize\n=\currenttypesize\advance\n by #1 \mathchoice
  {\setbox0\hbox{$\displaystyle-$} \w=.5\ht0\advance\w by-.5\dp0\setbox0
    \hbox{\typesize\n $\displaystyle-$} \advance\w by -.5\ht0\advance\w
    by .5\dp0\raise\w \hbox{\typesize\n$\displaystyle{\frac{#2}{#3}}$}}
  {\setbox0\hbox{$-$} \w=.5\ht0 \advance\w by -.5\dp0 \setbox0\hbox
    {\typesize\n $-$} \advance\w by-.5\ht0\advance\w by
    .5\dp0\raise\w\hbox{\typesize\n$\frac{#2}{#3}$}}
  {\setbox0\hbox{$\scriptstyle-$} \w=.5\ht0 \advance\w by-.5\dp0\setbox0
    \hbox{\typesize\n $\scriptstyle-$} \advance\w by -.5\ht0 \advance\w
    by .5\dp0 \raise\w\hbox{\typesize\n$\scriptstyle{\frac{#2}{#3}}$}}
  {\setbox0\hbox{$\scriptscriptstyle-$} \w=.5\ht0
    \advance\w by -.5\dp0 \setbox0\hbox{\typesize\n
    $\scriptscriptstyle-$} \advance\w by -.5\ht0 \advance\w by .5\dp0
    \raise\w\hbox{\typesize\n$\scriptscriptstyle{\frac{#2}{#3}}$}}  }

\def\d{{\rm d}}       

\def\Laplace{\bigtriangleup}  
\begin{document}

\title{Entropy production in phase field theories}
\author{P. V\'an$^{1,2}$ }
\address{$^1$Department of Theoretical Physics, Wigner Research Centre for Physics, H-1525 Budapest, Konkoly Thege Miklós u. 29-33., Hungary; //
and  $^2$Department of Energy Engineering, Faculty of Mechanical Engineering,  Budapest University of Technology and Economics, 1111 Budapest, Műegyetem rkp. 3., Hungary}
 
\date{\today}

\begin{abstract}
Allen-Cahn (Ginzburg-Landau) dynamics for scalar fields with heat conduction is treated in rigid bodies using a non-equilibrium thermodynamic framework with weakly nonlocal internal variables. The entropy production and entropy flux is calculated with the classical method of irreversible thermodynamics by separating full divergences.
\end{abstract}

\maketitle

\section{Introduction}

Phase field theories are dissipative. At least they seem to be dissipative, because the governing equations are parabolic and also there are various attempts to characterize the dissipation in the framework of their construction methods. However, they were originally introduced without thermodynamic considerations, with the help of a combination of variational and thermodynamic like methods \cite{CahHil58a,Cah61a,HohHal77a}. The universal background of these equations is questioned in spite of their widespread applicability and success in modelling various different phenomena \cite{HohKre15a,Emm08a}. Theoretically the most problematic aspect is their incompatibility with classical continuum theories. If they are dissipative, then the unification with classical theories, in particular heat conduction must be straightforward and important. Several conceptual frames were developed to understand this aspect, among them the most notable ones are the method of configurational forces \cite{Gur96a} and General Equation of Reversible and Irreversible Coupling (GenERIC) \cite{Ott05b}. Non-Equilibrium Thermodynamics with Internal Variables (NET-IV) is a natural framework to derive the governing differential equations of continua without variational considerations for both dissipative and nondissipative evolution \cite{Van18bc}. With dual internal variables one can model inertial effects, without or with dissipation \cite{VanAta08a,BerEta11a,BerEta11a1,BerVan17b,BerEta19a}. An  advantage of a pure thermodynamic background is the universality of the results: as long as the general conditions of the derivation are fulfilled, the results of the derivation are valid, independently of the micro or mesoscopic structure of the material. 

Thermodynamic compatibility of phase fields is a long time discussed and researched question, mostly with variational techniques. Then the identification of entropy flux and entropy production is problematic \cite{PenFif90a}. The relation to continuum balances is investigated mostly when the gradient extensions of classical fields are considered (see e.g. \cite{Sek11a}). In this short communication a single scalar internal variable  is treated with classical  thermodynamic methods in rigid heat conductors. Then Allen-Cahn type evolution is a solution of the entropy inequality. The relation of entropic and free energy representations are analysed. It is shown that heat flux is different with and without internal variables. 

\section{Classical variational-relaxational derivation of the Allen-Cahn equation}
\newcommand{\fs}{F}

What is the evolution equation of a single scalar field, an internal variable,  without any constraint in a continuum at rest? First one would look for a variational principle. However, then a second order time derivative and a nondissipative evolution cannot be avoided and we obtain something similar to continuum mechanics. There are several other systems, with diffusive properties and in this case the best evolution equations are obtained by a characteristic mixture of variational and thermodynamical ideas.

Let us denote the scalar field by  \(\alpha\). We assume, that the Helmholtz free energy density, $\fs$, depends on  this variable and its gradient: $\fs(\alpha,\nabla\alpha)$. For the sake of simplicity we consider the following square gradient form, a Ginzburg--Landau free energy function:
\eqn{GL_fun}{
	    \fs(\alpha, \nabla\alpha)= \fs_0(\alpha) + \frac{\gamma}{2}\nabla\alpha\cdot \nabla\alpha,
}

\noindent where  $\gamma$ is a nonnegative material parameter, which is scalar for isotropic continua. $\fs_0$ is the classical, local part of the free energy, that may be a double well potential, if $\alpha$ is an order parameter of a second order phase transition.  

Then, following the usual arguments, one assumes, that the rate of \(\alpha\)  in a body with volume \(V\) is negatively proportional to the change of the free energy, denoted by $\delta$:
\eqn{int_GL}{
\frac{d}{dt}\int_V \alpha {\rm d}V = -l \delta\int_V \fs(\alpha,\nabla\alpha)dV.
}

Assuming that this equality is valid for any \(V \) we obtain the general Allen--Cahn (Ginzburg--Landau) equation in the following form:
\eqn{GL_eq}{
\partial_t\alpha = -l   \frac{\delta \fs}{\delta \alpha} =
-l \big[\partial_\alpha \fs - \nabla\big(\partial_{\nabla\alpha} \fs\big)\big].
}

Here $\partial_t$ is the partial time derivative, $ \frac{\delta }{\delta \alpha}$ in the functional derivative, and $l$ is a material parameter. With the square gradient free energy, \re{GL_fun}, one arrives at the classical form of the equation:
\eqn{sGL_eq}{
\partial_t\alpha = 
-l \big(\partial_\alpha \fs_0 - \gamma\Laplace\alpha\big). 
}

The question is whether and in what sense the equation is dissipative and how is that related to the second law. In the following we will clarify this question and also show, that the Allen--Cahn equation follows from simple thermodynamics.

\section{Thermostatics of internal variables}

In our continuum theory the entropy is the function of the internal energy and also the scalar field $\alpha$ and its gradient. Therefore, the Gibbs relation for specific  quantities is written as
\eqn{sGrel}{
\d e = T\d s - \frac{A}{\rho}\d \alpha - \frac{{\mathbf A}}{\rho}\cdot \d \nabla\alpha.
}
Here $e$ is the specific internal energy, $s$ is the specific entropy, $T$ denotes the temperature,  $\rho$ is the density and $\nabla \alpha$ is the gradient of $\alpha$ field. The dot denotes the inner product of the corresponding vectors. The Gibbs relation is a convenient representation the specific entropy function $s(e,\alpha,\nabla\alpha)$, with  the partial derivatives:
\eqn{pardentr}{
\frac{\partial s}{\partial e} = \frac{1}{T}, \quad 
\frac{\partial s}{\partial \alpha} = \frac{A}{\rho T}, \quad
\frac{\partial s}{\partial \nabla\alpha} = \frac{{\mathbf A}}{\rho T}.
}

These partial derivatives define the internal variable related intensive quantities $A$ and $\mathbf A$. 

By means of the specific Helmholtz free energy, $f(T,\alpha,\nabla\alpha)$, considering the definition by Legendre transformation, $f= e- Ts$, one can see easily that 
\eqn{pardf}{
\frac{\partial f}{\partial T} = -s, \quad 
\frac{\partial f}{\partial \alpha} = -\frac{A}{\rho}, \quad
\frac{\partial f}{\partial \nabla\alpha} = -\frac{{\mathbf A}}{\rho}.
}

In case of a continuum at rest the specific free energy is related to the free energz density as $\fs = \rho f$.  With these expressions the thermostatics for a classical field theory with a single scalar weakly nonlocal internal variable is defined. The derivation of the corresponding relations for densities and also with global quantities is straightforward, introducing the extensivity of the thermodynamic potentials \cite{BerVan17b}.

\section{Entropy production}

The substantial balance of internal energy is given as
\eqn{intebal}{
\rho \dot e + \nabla\cdot {\mathbf q} = 0,
}
where {\bf q} is the heat flux, the overdot  denotes substantial time differentiation and $\nabla\cdot $ denotes divergence. In our case, for rigid heat conductors, the substantial time derivative is equal to the partial one. For the calculation of the entropy balance we follow the classical method of de Groot and Mazur with the identification of the  entropy flux by a convenient separation of full divergences \cite{GroMaz62b}. The method was applied by Maugin for internal variables, too \cite{Mau06a}. Therefore, for the time derivative of the entropy we obtain, that 
\eqn{entrbal0}{
\rho \dot s(e,\alpha,\nabla\alpha) = -\frac{1}{T}\nabla{\mathbf q} + \frac{A}{T}\dot \alpha +  \frac{\mathbf A}{T}\cdot\frac{\d }{\d t}(\nabla\alpha),
}
where we have substituted the time derivative of the internal energy with \re{intebal} and represented the overdot notation of the substantial time derivative by $\frac{\d }{\d t}$. In case of rigid bodies the spatial and substantial derivatives commute, therefore $\frac{\d }{\d t} \nabla = \nabla \frac{\d }{\d t}$. In this case, continuing the calculation
\eqn{entrbal1}{
\rho \dot s= -\nabla\cdot\frac{\mathbf q}{T} + {\mathbf q}\cdot \nabla\frac{1}{T} + \frac{A}{T}\dot \alpha +  \nabla\cdot\left(\frac{\mathbf A \dot\alpha}{T}\right) -\dot\alpha \nabla\cdot\frac{\mathbf A}{T}.
}

Therefore the complete entropy balance can be written in the following form:
\eqn{entrbal}{
\rho \dot s + \nabla\cdot\left(\frac{{\mathbf q} -{\mathbf A}\dot \alpha}{T}\right) = ({\mathbf q} -{\mathbf A}\dot \alpha)\cdot\nabla\frac{1}{T} + 
		\frac{\dot \alpha}{T}\left(A- \nabla\cdot\dot {\mathbf A}\right) \geq 0.
}

Here we can identify the entropy flux $\mathbf J$ and the entropy production $\sigma$ as follows:
\eqn{sflu}{
{\mathbf J} &= \frac{{\mathbf q} -{\mathbf A}\dot \alpha}{T} \nl{sprod}
\sigma &= ({\mathbf q} -{\mathbf A}\dot \alpha)\cdot\nabla\frac{1}{T} + 
		\frac{\dot \alpha}{T}\left(A- \nabla\cdot \dot {\mathbf A}\right) \geq 0.
}

The constitutive relations can be determined by solving the inequality \re{sprod} and recognizing that it leads to the evolution equation of $\alpha$, and the constitutive function of the heat flux $\mathbf q$. In case of linear relationship between the thermodynamic fluxes and forces and for isotropic materials one obtains, that 
\eqn{tr1}{
\dot \alpha = l\left( A- \nabla\cdot {\mathbf A} \right) &= 
	l\rho T\left(\frac{\partial s}{\partial \alpha} - \nabla\cdot \frac{\partial s}{\partial(\nabla\alpha)}\right) =
	-l\rho\left(\frac{\partial f}{\partial \alpha} - \nabla\cdot \frac{\partial f}{\partial(\nabla\alpha)}\right), \nl{tr2}
{\mathbf q} = \dot\alpha {\mathbf A} + \lambda\nabla\frac{1}{T} &= 
\rho T\dot \alpha \frac{\partial s}{\partial(\nabla\alpha)}  + \lambda\nabla\frac{1}{T} = 
-\rho\dot \alpha \frac{\partial f}{\partial(\nabla\alpha)}  + \lambda\nabla\frac{1}{T}.
} 

Here $\lambda = \lambda_F T^2$, where $\lambda_F$ is the Fourier heat conduction coefficient. We can recognize that \re{tr1} is the Cahn-Allen-Ginzburg-Landau equation. The second law requires nonnegative relaxation and heat conduction coefficients $l$ and $\lambda_F$. With a given thermodynamic potential ($s$, or $f$), \re{intebal} and \re{tr1}-\re{tr2} are a closed system of differential equations to be solved for considering phase fields coupled to thermal effects.

\section{Discussion}

The derivation can be extended easily with mechanical interaction for fluids and elastic solids. Then the restriction of constant density must be released with the present thermostatic representation of the internal variable and its gradient \cite{BerVan17b}. 

It is also remarkable that the simple and heuristic exploitation method fo separation of divergences is compatible with the more rigorous Liu or Coleman-Noll procedures in as it was shown in \cite{Van18bc}. It was also shown that for a Cahn-Hilliard evolution the rigorous exploitation is more technical.

For phase transition models the concavity of the entropy or/and the proper convexity relations for free energy are important requirements. This property ensures the stability of equilibria and the basin of attraction is related to simply connected concave regions. With additional considerations for boundary conditions the total entropy is a good candidate for a Ljapunov functional of equilibria as it is indicated in \cite{PenFif90a}.

\section{Acknowledgement}   
The work was supported by the grants National Research, Development and Innovation Office - NKFIH 116197(116375), NKFIH 124366(124508) and NKFIH 123815. 

The paper is dedicated to Jüri Engelbrecht on the occasion of his 80th birthday.  

\bibliographystyle{spbasic}

\end{document}